\documentclass{article}
\usepackage{latexsym}
\usepackage{amsmath}
\usepackage{amsfonts}
\usepackage{amssymb}
\usepackage{graphicx}
\newtheorem{theorem}{Theorem}

\begin{document}

\author{Brian C. Hall\\Department of Mathematics\\University of Notre Dame\\Notre Dame, IN 46556 U.S.A.\\bhall@nd.edu}
\title{Coherent states and the quantization of 1+1-dimensional Yang-Mills theory}
\date{December, 2000}
\maketitle
\begin{abstract}
This paper discusses the canonical quantization of 1+1-dimensional Yang-Mills
theory on a spacetime cylinder, from the point of view of coherent states, or
equivalently, the Segal-Bargmann transform. Before gauge symmetry is imposed,
the coherent states are simply ordinary coherent states labeled by points in
an infinite-dimensional linear phase space. Gauge symmetry is imposed by
projecting the original coherent states onto the gauge-invariant subspace,
using a suitable regularization procedure. We obtain in this way a new family
of ``reduced'' coherent states labeled by points in the reduced phase space,
which in this case is simply the cotangent bundle of the structure group $K.$
The main result explained here, obtained originally in a joint work of the
author with B. Driver, is this: The reduced coherent states are precisely
those associated to the generalized Segal-Bargmann transform for $K,$ as
introduced by the author from a different point of view. This result agrees
with that of K. Wren, who uses a different method of implementing the gauge
symmetry. The coherent states also provide a rigorous way of making sense out
of the quantum Hamiltonian for the unreduced system. Various related issues
are discussed, including the complex structure on the reduced phase space and
the question of whether quantization commutes with reduction.
\end{abstract}
\tableofcontents

\section{Introduction}

The quantization of Yang-Mills theory is an important example of the
quantization of reduced Hamiltonian systems. This paper concerns the simplest
non-trivial case of quantized Yang-Mills theory, namely, pure Yang-Mills on a
spacetime cylinder. The main result described here is from a joint work
\cite{DH1} with B. Driver. However, I also discuss a number of related
conceptual points, and the emphasis here is on the ideas rather than the
mathematical technicalities.

Driver and I use as our main tool the Segal--Bargmann transform, or
equivalently, coherent states. We reach two main conclusions. First, upon
reduction the \textit{ordinary} coherent states on the space of connections
become the \textit{generalized} coherent states in the sense of \cite{H1} on
the finite-dimensional compact structure group. Second, coherent states
provide a way to make rigorous the generally accepted idea that upon reduction
the Laplacian for the infinite-dimensional space of connections becomes the
Laplacian on the structure group. In the rest of the introduction I give a
schematic description of the paper. More details are found in the body of the
paper and in \cite{DH1}. See also \cite{H7} for additional exposition.

Driver and I use the canonical quantization approach rather than the
path-integral approach, and we work in the temporal gauge. As stated, we
assume that spacetime is a cylinder, namely, $S^{1}\times\mathbb{R}.$ We fix a
compact connected structure group $K,$ which I will assume here is simple
connected, with Lie algebra $\frak{k}.$ The configuration space for the
classical theory is the space $\mathcal{A}$ of $\frak{k}$-valued connection
1-forms over the spatial circle. The gauge group $\mathcal{G}$, consisting of
maps of the spatial circle into $K,$ acts naturally on $\mathcal{A}.$ The
based gauge group $\mathcal{G}_{0},$ consisting of gauge transformations that
equal the identity at one fixed point in the spatial circle, acts freely on
$\mathcal{A},$ and the quotient $\mathcal{A}/\mathcal{G}_{0}$ is simply the
compact structure group $K.$ This reflects that in this simple case the only
gauge-invariant quantity is the holonomy of a connection around the spatial circle.

Meanwhile we have the complexification of $\mathcal{A},$ namely,
$\mathcal{A}_{\mathbb{C}}:=\mathcal{A}+i\mathcal{A},$ which is identifiable
with the cotangent bundle of $\mathcal{A}$ and is the phase space for the
unreduced system. We have also the complexification $K_{\mathbb{C}}$ of the
structure group $K,$ which is identifiable with the cotangent bundle of $K.$
Here $K_{\mathbb{C}}$ is the unique simply connected complex Lie group whose
Lie algebra is $\frak{k}+i\frak{k}.$ One defines in the obvious way the based
complexified gauge group $\mathcal{G}_{0,\mathbb{C}}$, which acts
holomorphically on $\mathcal{A}_{\mathbb{C}}.$ The quotient $\mathcal{A}%
_{\mathbb{C}}/\mathcal{G}_{0,\mathbb{C}}$ is $K_{\mathbb{C}}.$ This is the
reduced phase space for the theory.

Now we have the ordinary Segal--Bargmann transform for $\mathcal{A}$, which
maps from an $L^{2}$ space of functions on $\mathcal{A}$ to an $L^{2}$ space
of \textit{holomorphic} functions on $\mathcal{A}_{\mathbb{C}}.$ Much more
recently there is a generalized Segal--Bargmann transform for $K$ \cite{H1},
which maps from an $L^{2}$ space of functions on $K$ to an $L^{2}$ space of
holomorphic functions on $K_{\mathbb{C}}.$ The gist of \cite{DH1} is that the
\textit{ordinary} Segal--Bargmann transform for $\mathcal{A},$ when restricted
to the gauge-invariant subspace is precisely the \textit{generalized}
Segal--Bargmann transform for $\mathcal{A}/\mathcal{G}_{0}=K.$ To say the same
thing in the language of coherent states, taking the ordinary coherent states
for $\mathcal{A}$ and projecting them onto the gauge-invariant subspace gives
the generalized coherent states for $K,$ in the sense of \cite{H1}. So
\cite{DH1} gives a new way of understanding the generalized Segal--Bargmann
transform (or generalized coherent states) for a compact Lie group $K.$

Another purpose for \cite{DH1} is to understand the Hamiltonian for the
Yang-Mills theory, which at the unreduced level is a multiple of the Laplacian
$\Delta_{\mathcal{A}}$ for $\mathcal{A}.$ (The usual curvature term is zero in
this case, since there cannot be any curvature on the one-dimensional space
manifold $S^{1}.$) The difficulty lies in making sense of $\Delta
_{\mathcal{A}}$ as a reasonable operator in the quantum Hilbert space.
However, the Segal--Bargmann transform for $\mathcal{A}$ is expressible in
terms of $\Delta_{\mathcal{A}},$ and it is well-defined. The Segal--Bargmann
transform for $K$ is expressed in a precisely parallel way in terms of the
Laplacian for $K.$ Theorem 5.2 of \cite{DH1} (see Theorem \ref{sb4} below)
states that the Segal--Bargmann transform for $\mathcal{A}$ becomes the
generalized Segal--Bargmann transform for $K$ when restricted to the
gauge-invariant subspace. This is formally equivalent to the following
generally accepted principle.
\begin{equation}
\text{On the gauge-invariant subspace},\text{ }\Delta_{\mathcal{A}}\text{
reduces to }\Delta_{K}.\label{main.idea}%
\end{equation}
Driver and I wish to interpret Theorem 5.2 of \cite{DH1} as a rigorous version
of this principle, which does not make mathematical sense as written. (See
Section 3.) Thus the Hamiltonian for the reduced system becomes a multiple of
$\Delta_{K}.$

I discuss three additional points. First, I consider the question of finding
the ``right'' complex structure on the reduced phase space $T^{\ast}\left(
K\right)  .$ Although having such a complex structure is necessary in order to
construct a Segal--Bargmann transform, it is not \textit{a priori} obvious
what the correct complex structure is. I explain in Section 6 how a complex
structure on the reduced phase space arises naturally out of the reduction
process, and show that this complex structure is the same as the one
previously considered at an ``intrinsic'' level. Second, I discuss why, even
at a formal level, $\Delta_{\mathcal{A}}$ \textit{should} go to $\Delta_{K}$
on the invariant subspace. For Yang-Mills theory in higher dimensions, K.
Gaw\c{e}dzki \cite{Ga} has shown that the reduced and the unreduced Laplacians
do not agree (even at a formal level) when applied to gauge-invariant
functions. So there is something geometrically special about the
1+1-dimensional case, as discussed in Section 7. Finally, I consider the
possibility of doing things in the opposite order, namely, first passing to
the reduced phase space $K_{\mathbb{C}},$ and then constructing coherent
states by means of geometric quantization. It turns out that the two
procedures give the same answer, \textit{provided} that on includes as part of
the geometric quantization the ``half-form correction.'' Thus one may say that
in this case, ``quantization commutes with reduction.'' It is unlikely that
such a result holds (even formally) for higher-dimensional Yang-Mills theory.

I have tried to emphasize the concepts rather than the mathematical
technicalities. Some of the subtleties that I have glossed over elsewhere are
discussed in Section 9.

\textit{Acknowledgments.} The idea of deriving the generalized Segal--Bargmann
transform from the infinite-dimensional ordinary Segal--Bargmann transform is
due to L. Gross and P. Malliavin \cite{GM}. However, \cite{GM} is a paper on
stochastic analysis, and it was not intended to be about Yang-Mills theory.
What I am here calling the gauge group $\mathcal{G}_{0}$ they call the loop
group, and its action in \cite{GM} is not unitary. To apply the approach of
Gross and Malliavin in the Yang-Mills setting, Driver and I modified that
approach so as to make the action of $\mathcal{G}_{0}$ unitary. (More
precisely, we take a certain limit under which the action of $\mathcal{G}_{0}$
becomes formally unitary.)

The idea that the generalized coherent states for $K$ could be obtained from
the ordinary coherent states for $\mathcal{A}$ by reduction is due to K. Wren
\cite{W}. Wren uses the ``Rieffel induction'' approach proposed by N. Landsman
\cite{L1} and carried out in the abelian case by Landsman and Wren \cite{LW}.
See also the exposition in the book of Landsman \cite[Sect. IV.3.7]{L2}. I
describe in Section 5 the relationship of our results to those of Wren.

I am indebted to Bruce Driver for clarifying to me many aspects of what is
discussed here. I also acknowledge valuable discussions with Andrew Dancer,
and I thank Dan Freed for a valuable suggestion regarding the half-form correction.

\section{Classical Yang-Mills theory on a spacetime cylinder}

Yang-Mills theory on a spacetime cylinder is an exactly solvable model. (See
for example \cite{R}.) Nevertheless, I believe that there are things to learn
here, both classically and quantum mechanically, by comparing what happens
before gauge symmetry is imposed to what happens afterward. I begin with the
classical theory, borrowing heavily from the treatment of Landsman \cite[Sect.
IV.3.6]{L2}.

We work on the spacetime manifold $S^{1}\times\mathbb{R},$ with $S^{1}$ being
space and $\mathbb{R}$ time. Fix a connected compact Lie group $K$, the
structure group, which for simplicity I take to be simply connected, and fix
an Ad-invariant inner product on the Lie algebra $\frak{k}$ of $K$. We work in
the temporal gauge, which has the advantage of allowing the classical
Yang-Mills equations to be put into Hamiltonian form. The temporal gauge is
only a partial gauge-fixing, leaving still a large gauge group $\mathcal{G},$
namely the group of mappings of the space manifold $S^{1}$ into the structure
group $K.$ Note that the gauge group is just a loop group in this case. I will
concern myself only with the based gauge group $\mathcal{G}_{0}, $ consisting
of maps of $S^{1}$ into $K$ that equal the identity at one fixed point in
$S^{1}.$ This group acts freely on $\mathcal{A}$. The remaining gauge symmetry
can easily be added later.

In the temporal gauge, the Yang-Mills equations have a configuration space
$\mathcal{A}$ consisting of connections on the space manifold. The connections
are 1-forms with values in the Lie algebra $\frak{k}.$ Since our space
manifold is one-dimensional, we may think of the connections as $\frak{k}%
$-valued functions. There is a natural norm on $\mathcal{A}$ given by
\[
\left\|  A\right\|  ^{2}=\int_{0}^{1}\left|  A\left(  \tau\right)  \right|
^{2}\,d\tau,\quad A\in\mathcal{A}.
\]
Here $S^{1}$ is the interval $\left[  0,1\right]  $ with ends identified, and
$\left|  A\left(  \tau\right)  \right|  ^{2}$ is computed using the inner
product on $\frak{k}.$ The norm allows us to define a distance function
\[
d\left(  A,B\right)  :=\left\|  A-B\right\|  .
\]
The gauge group $\mathcal{G}_{0}$ acts on $\mathcal{A}$ by
\begin{equation}
\left(  g\cdot A\right)  _{\tau}=g_{\tau}A_{\tau}g_{\tau}^{-1}-\frac{dg}%
{d\tau}g_{\tau}^{-1}.\label{action}%
\end{equation}
The map $A\rightarrow gAg^{-1}$ is linear, invertible, and norm-preserving,
hence a ``rotation'' of $\mathcal{A}.$ The map $A\rightarrow g\cdot A$ is
affine and satisfies
\[
d\left(  g\cdot A,g\cdot B\right)  =d\left(  A,B\right)  .
\]
So a gauge transformation is a combination of a rotation and a translation of
$\mathcal{A}.$

The phase space of the theory is the cotangent bundle of $\mathcal{A},$
$T^{\ast}\!\left(  \mathcal{A}\right)  \cong\mathcal{A}+\mathcal{A}.$ The
action of $\mathcal{G}_{0}$ on $\mathcal{A}$ extends in a natural way to an
action on $\mathcal{A}+\mathcal{A}$ given by
\[
g\cdot\left(  A,P\right)  =\left(  g\cdot A,gPg^{-1}\right)  .
\]
Note that the translation part of (\ref{action}) affects only the ``position''
$A$ and not the ``momentum'' $P.$

The Yang-Mills equations take place in the phase space $\mathcal{A}%
+\mathcal{A}$ and have three parts. First we have a dynamical part. The
equations of motion are just Hamilton's equations, for the Hamiltonian
function
\[
H\left(  A,P\right)  =\frac{1}{2}\left\|  P\right\|  ^{2}.
\]
Normally there would be another term involving the curvature of $A,$ but that
term is necessarily zero in this case, since $S^{1}$ is one-dimensional. Thus
the solutions of Hamilton's equations are embarrassingly easy to write down:
the general solution is
\[
\left(  A\left(  t\right)  ,P\left(  t\right)  \right)  =\left(  A_{0}%
+tP_{0},P_{0}\right)  .
\]
This is just free motion in $\mathcal{A}.$ Observe that the Hamiltonian $H$ is
invariant under the action of $\mathcal{G}_{0}$ on $\mathcal{A}+\mathcal{A}.$

Second we have a constraint part. This says that the solutions (trajectories
in $\mathcal{A}+\mathcal{A}$) have to lie in a certain set, which I will
denote $J^{-1}\left(  0\right)  ,$ which is ``the zero set of the moment
mapping for the action of $\mathcal{G}_{0}.$'' I will not repeat here the
formulas, which may be found for example in \cite[Sect. 2]{DH1} or \cite[Sect.
IV.3.6]{L2}. This constraint is of a simple sort, in that $J^{-1}\left(
0\right)  $ is invariant under the dynamics and under the action of
$\mathcal{G}_{0}$ on $\mathcal{A}+\mathcal{A}.$ So the constraint does not
alter the dynamics, it merely restricts us to certain special solutions of the
original equations of motion.

Third we have a philosophical part. This says that the only functions on phase
space that are physically observable are the gauge-invariant ones.

The last two points together say that we may think of the dynamics as taking
place in $J^{-1}\left(  0\right)  /\mathcal{G}_{0},$ which is the same as
$T^{\ast}\!\left(  \mathcal{A}/\mathcal{G}_{0}\right)  .$ The process of
restricting to $J^{-1}\left(  0\right)  $ and then dividing by the action of
$\mathcal{G}_{0}$ is called Marsden--Weinstein or symplectic reduction. Since
the Hamiltonian function $H$ is $\mathcal{G}_{0}$-invariant, it makes sense as
a function on $J^{-1}\left(  0\right)  /\mathcal{G}_{0}.$

Now, we are in a very simple situation, with the space manifold being just a
circle. In this case two connections are gauge-equivalent if and only if they
have the same holonomy around the spatial circle. So the orbits of
$\mathcal{G}_{0}$ are labeled by the holonomy $h\left(  A\right)  $ of a
connection $A$ around the circle, where for $A\in\mathcal{A},$ $h\left(
A\right)  $ is an element of the structure group $K.$ It is easily seen that
any $x\in K$ can be the holonomy of some $A,$ and so we have
\[
\mathcal{A}/\mathcal{G}_{0}\cong K.
\]
Thus
\[
J^{-1}\left(  0\right)  /\mathcal{G}_{0}\cong T^{\ast}\!(\mathcal{A}%
/\mathcal{G}_{0})\cong T^{\ast}\!(K).
\]
After the reduction, the dynamics become geodesic motion in $K,$ where
explicitly the geodesics in $K$ may be written as $\gamma\left(  t\right)
=xe^{tX}$with $x\in K$ and $X\in\frak{k}.$

We require one last discussion before turning to the quantum theory. We may
think of $\mathcal{A}+\mathcal{A}$ as the complex vector space $\mathcal{A}%
_{\mathbb{C}}=\mathcal{A}+i\mathcal{A},$ in the same way that we think of
$T^{\ast}\!(\mathbb{R})\cong\mathbb{R+R}$ as $\mathbb{C}.$ We may then think
of elements of $\mathcal{A}_{\mathbb{C}}$ as functions (or 1-forms) with
values in the complexified Lie algebra $\frak{k}_{\mathbb{C}}=\frak{k}%
+i\frak{k}.$ The action of $\mathcal{G}_{0}$ extends to an action on
$\mathcal{A}_{\mathbb{C}}$ by
\[
\left(  g\cdot Z\right)  _{\tau}=g_{\tau}Z_{\tau}g_{\tau}^{-1}-\frac{dg}%
{d\tau}g_{\tau}^{-1},
\]
where $Z:\left[  0,1\right]  \rightarrow\frak{k}_{\mathbb{C}}.$ Note that the
translation part in the real direction; that is, $\frac{dg}{d\tau}g_{\tau
}^{-1}$ is in $\mathcal{A}.$ One can think of elements of $\mathcal{A}%
_{\mathbb{C}}$ as complex connections and thus define their holonomy. But the
holonomy now takes values in the \textit{complexified} group $K_{\mathbb{C}},$
where $K_{\mathbb{C}}$ is the unique simply connected complex Lie group with
Lie algebra $\frak{k}+i\frak{k}.$ (For example, if $K=SU(n)$ then
$K_{\mathbb{C}}=SL(n;\mathbb{C}).$) The complexified based gauge group
$\mathcal{G}_{0,\mathbb{C}}$ is then the group of based loops with values in
$K_{\mathbb{C}}.$ The same reasoning as on $\mathcal{A}$ shows that the only
$\mathcal{G}_{0,\mathbb{C}}$-invariant quantity on $\mathcal{A}_{\mathbb{C}}$
is the holonomy; so $\mathcal{A}_{\mathbb{C}}/\mathcal{G}_{0,\mathbb{C}%
}=K_{\mathbb{C}}.$

It turns out that restricting to the zero set of the moment mapping and then
dividing out by the action of $\mathcal{G}_{0}$ gives the same result as
working on the whole phase space and then dividing out by the action of
$\mathcal{G}_{0,\mathbb{C}}.$ Thus
\[
J^{-1}\left(  0\right)  /\mathcal{G}_{0}=\mathcal{A}_{\mathbb{C}}%
/\mathcal{G}_{0,\mathbb{C}}=K_{\mathbb{C}}.
\]
On the other hand, we have already said that $J^{-1}\left(  0\right)
/\mathcal{G}_{0}$ is identifiable with $T^{\ast}\!(K).$ So we have a natural
identification
\[
K_{\mathbb{C}}\cong T^{\ast}\!(K).
\]
This is explained in detail in Section 6 and the resulting identification is
given there explicitly.

\section{Formal and semiformal quantization}

In this section we will see what is involved in trying to quantize this
system. This discussion will set the stage for the entrance of the
Segal--Bargmann transform and the coherent states in the next two sections.

Let us first try to quantize our Yang-Mills example at a purely formal level,
that is, without worrying too much whether our formulas make sense. I want to
do the quantization \textit{before} the reduction by $\mathcal{G}_{0}.$ If we
did the reduction before the quantization, then we would have a
finite-dimensional system, which is easily quantized. So it is of interest to
do the quantization first and see if this gives the same result. See \cite{R},
where quantization is done after the reduction, and \cite{Di}, where
quantization is done before the reduction.

Since our system has a configuration space $\mathcal{A},$ we may formally take
our unreduced quantum Hilbert space to be
\[
L^{2}\left(  \mathcal{A},\mathcal{D}A\right)  ,
\]
where $\mathcal{D}A$ is the fictitious Lebesgue measure on $\mathcal{A}.$ The
quantization of the constraint equation (see \cite[Sect. 2]{DH1}) then imposes
the condition that our ``wave functions'' be $\mathcal{G}_{0}$-invariant. Note
that the quantization of the second part of the classical theory (the
constraint) automatically incorporates the third part as well (the
$\mathcal{G}_{0}$-invariance). So we want the reduced (physical) quantum
Hilbert space to be
\[
L^{2}\left(  \mathcal{A},\mathcal{D}A\right)  ^{\mathcal{G}_{0}}:=\left\{
f\in L^{2}\left(  \mathcal{A},\mathcal{D}A\right)  |f\left(  g\cdot A\right)
=f\left(  A\right)  ,\forall g,A\right\}  .
\]

Recall that in our example, in which space is a circle, two connections are
$\mathcal{G}_{0}$-equivalent if and only if they have the same holonomy around
the spatial circle. That means that a $\mathcal{G}_{0}$-invariant function
must be of the form
\begin{equation}
f\left(  A\right)  =\phi\left(  h\left(  A\right)  \right)  ,\label{inv.form}%
\end{equation}
where $h\left(  A\right)  \in$ $K$ is the holonomy of $A$ and where $\phi$ is
a function on the structure group $K.$ Furthermore, as we shall see more
clearly in the next section, it is reasonable to think that for a function of
the form (\ref{inv.form}), integrating $\left|  f\left(  A\right)  \right|
^{2}$ with respect to $\mathcal{D}A$ is the same as integrating $\left|
\phi\left(  g\right)  \right|  ^{2}$ with respect to a multiple of the Haar
measure on $K.$ Thus (formally)
\begin{equation}
L^{2}\left(  \mathcal{A},\mathcal{D}A\right)  ^{\mathcal{G}_{0}}\cong
L^{2}\left(  K,C\cdot dg\right)  ,\label{inv.space}%
\end{equation}
for some (infinite) constant $C.$ This is our physical Hilbert space.

Next we consider the Hamiltonian. Formally quantizing the function $\frac
{1}{2}\left\|  P\right\|  ^{2}$ in the usual way gives
\begin{equation}
\hat{H}=-\frac{\hbar^{2}}{2}\Delta_{\mathcal{A}}=-\frac{\hbar^{2}}{2}%
\sum_{k=1}^{\infty}\frac{\partial^{2}}{\partial x_{k}^{2}},\label{lap.def}%
\end{equation}
where the $x_{k}$'s are coordinates with respect to an orthonormal basis of
$\mathcal{A}.$ We must now try to determine how $\hat{H}$ acts on the
$\mathcal{G}_{0}$-invariant subspace. In light of what happens when performing
the reduction before the quantization, it is reasonable to guess that on the
invariant subspace $\Delta_{\mathcal{A}}$ reduces to $\Delta_{K},$ that is,
\begin{equation}
\Delta_{\mathcal{A}}\left[  \phi\left(  h\left(  A\right)  \right)  \right]
=\left(  \Delta_{K}\phi\right)  \left(  h\left(  A\right)  \right)
.\label{lap.eq}%
\end{equation}
(See also \cite{Di,W}. See Section 7 for a discussion of why (\ref{lap.eq}) is
formally correct.) If we accept this and if we ignore the infinite constant
$C$ in (\ref{inv.space}) then we conclude that our quantum Hilbert space is
\[
L^{2}\left(  K,dx\right)
\]
and our Hamiltonian is
\[
\hat{H}=-\frac{\hbar^{2}}{2}\Delta_{K}.
\]
This concludes the formal quantization of our system.

We now begin to consider how to make this mathematically precise. One approach
is to forget about the measure theory (i.e. the Hilbert space) and to try to
prove (\ref{lap.eq}). As it turns out, the answer is basis-dependent--choosing
different bases in (\ref{lap.def}) will give different answers. Another way of
saying this is that the matrix of second derivatives of a function $f$ of the
form (\ref{inv.form}) is in general not of trace class. However, if one uses
the most obvious sort of basis, then indeed it turns out that (\ref{lap.def})
is true. See the appendix of \cite{DH1}.

Even without the problem of basis-dependence, the above approach is
unsatisfying because we would like to define $\hat{H}$ as an operator in some
Hilbert space. Since Lebesgue measure $\mathcal{D}A$ does not actually exist,
one reasonable procedure is to ``approximate'' $\mathcal{D}A$ by a Gaussian
measure $dP_{s}\left(  A\right)  $ with large variance $s.$ This means that
$P_{s}$ is formally given by the expression
\[
dP_{s}\left(  A\right)  =c_{s}e^{-\left\|  A\right\|  ^{2}/2s}\mathcal{D}A,
\]
where $c_{s}$ is supposed to be a normalization constant that makes the total
integral one. Formally as $s\rightarrow\infty$ we get back a multiple of
Lebesgue measure $\mathcal{D}A.$ The measure $P_{s}$ does exist rigorously,
provided that one allows sufficiently non-smooth connections.

There is good news and bad news about this approach. First the good news. 1)
Even though the connections in the support of $P_{s}$ are not smooth, the
holonomy of such a connection makes sense, as the solution to a
\textit{stochastic} differential equation. 2) If we define the gauge-invariant
subspace to be
\[
L^{2}\left(  \mathcal{A},P_{s}\right)  ^{\mathcal{G}_{0}}=\left\{  f\text{
}|\,\text{for all }g\in\mathcal{G}_{0},\text{ }f\left(  g\cdot A\right)
=f\left(  A\right)  \text{ for }P_{s}\text{-almost every }A\right\}  ,
\]
then the Gross ergodicity theorem \cite{G2} asserts that $L^{2}\left(
\mathcal{A},P_{s}\right)  ^{\mathcal{G}_{0}}$ is precisely what we expect,
namely, the space of functions of the form $f\left(  A\right)  =\phi\left(
h\left(  A\right)  \right)  ,$ with $\phi$ a function on $K.$ 3) There is a
natural dense subspace of $L^{2}\left(  \mathcal{A},P_{s}\right)  $ on which
$\Delta_{\mathcal{A}}$ is unambiguously defined, consisting of smooth cylinder
functions. Here a cylinder function is one which depends on only finitely many
of the infinitely many variables in $\mathcal{A}.$ See \cite[Defn. 4.2]{DH1}.

Note that the map which takes $f\left(  A\right)  $ to $f\left(  g\cdot
A\right)  $ is not unitary, because the measure $P_{s}$ is not invariant under
the action of $\mathcal{G}_{0}.$ Driver and I wish to avoid ``unitarizing''
the action of $\mathcal{G}_{0},$ because if we did unitarize then there would
be no gauge-invariant subspace. (See \cite{DH2}.) Instead of unitarizing the
action for a fixed value of $s,$ we will eventually let $s $ tend to infinity,
at which point unitarity will be formally recovered.

The bad news about this approach is that $\Delta_{\mathcal{A}}$ is not a
closable operator, and that functions of the holonomy are not cylinder
functions. The non-closability of $\Delta_{\mathcal{A}}$ means that if we
approximate $\phi\left(  h\left(  A\right)  \right)  $ by cylinder functions,
then the value of $\Delta_{\mathcal{A}}\phi\left(  h\left(  A\right)  \right)
$ depends on the choice of approximating sequence. So we still have a major
problem in making mathematical sense out of the Hamiltonian in our quantum theory.

\section{The Segal--Bargmann transform to the rescue}

In this section I will explain how the Segal--Bargmann transform can be used
to make sense out of the quantization of the Hamiltonian. At the same time, we
will see how the generalized Segal--Bargmann transform for the structure group
$K$ arises from the restriction of the ordinary Segal--Bargmann transform for
the gauge-invariant subspace. Although it is technically easier to describe
the quantization in terms of the Segal--Bargmann transform, there is a
formally equivalent description in terms of coherent states, as I will explain
in the next section. See \cite{B,S1,S2,S3} and also \cite{BSZ,H6} for results
on the ordinary Segal--Bargmann transform.

Let me explain the normalization of the Segal--Bargmann transform that I wish
to use, first for the finite-dimensional space $\mathbb{R}^{d}.$ Let
$\mathcal{H}(\mathbb{C}^{d})$ denote the space of holomorphic (complex
analytic) functions on $\mathbb{C}^{d}.$ For any positive constant $\hbar,$
define
\[
C_{\hbar}:L^{2}(\mathbb{R}^{d},dx)\rightarrow\mathcal{H}(\mathbb{C}^{d})
\]
by the formula
\begin{equation}
C_{\hbar}f\left(  z\right)  =\left(  2\pi\hbar\right)  ^{-d/2}\int
_{\mathbb{R}^{d}}e^{-\left(  z-x\right)  ^{2}/2\hbar}f\left(  x\right)
\,dx,\quad z\in\mathbb{C}^{d}.\label{ct.form}%
\end{equation}
Here $\left(  z-x\right)  ^{2}$ means $\Sigma\left(  z_{k}-x_{k}\right)
^{2}.$ If we restrict attention to $z\in\mathbb{R}^{d},$ then this is the
standard expression for the solution of the heat equation $\partial
u/\partial\hbar=\frac{1}{2}\Delta u,$ at time $\hbar$ and with initial
condition $f.$ Thus we may write
\[
C_{\hbar}f=\text{ analytic continuation of }e^{\hbar\Delta/2}f.
\]
Here $e^{\hbar\Delta/2}f$ is just a mnemonic for the solution of the heat
equation with initial condition $f,$ and the analytic continuation is in the
space variable (analytic continuation from $\mathbb{R}^{d}$ to $\mathbb{C}%
^{d}$). Because $\hbar$ is playing the role of time in the heat equation, it
is tempting call this parameter $t$ instead of $\hbar;$ this is what we do in
\cite{DH1}.

Now let $\nu_{\hbar}$ be the measure on $\mathbb{C}^{d}$ given by
\[
d\nu_{\hbar}\left(  z\right)  =\left(  \pi\hbar\right)  ^{-d/2}e^{-\left(
\operatorname{Im}z\right)  ^{2}/\hbar}dz,
\]
where $dz$ refers to the $2d$-dimensional Lebesgue measure on $\mathbb{C}^{d}.$

\begin{theorem}
[Segal-Bargmann transform]For each positive value of $\hbar,$ $C_{\hbar}$ is a
unitary map of $L^{2}(\mathbb{R}^{d},dx)$ onto $\mathcal{H}L^{2}%
(\mathbb{C}^{d},\nu_{\hbar}),$ where $\mathcal{H}L^{2}$ denotes the space of
entire holomorphic functions on $\mathbb{C}^{d}$ which are square-integrable
with respect to $\nu_{\hbar}.$
\end{theorem}

This is not quite the form of the transform given by either Segal or Bargmann.
Comparing to Bargmann's map $A$ (and taking $\hbar=1$ since that is what
Bargmann does) we have
\[
C_{1}f\left(  z\right)  =\left(  4\pi\right)  ^{-d/4}e^{-z^{2}/4}Af\left(
\frac{z}{\sqrt{2}}\right)  .
\]
The factor of $\sqrt{2}$ accounts for the difference between Bargmann's
convention that $z=\left(  x+ip\right)  /\sqrt{2}$ and my convention that
$z=x+ip,$ which is preferable for me because on a more general manifold, the
map $z\rightarrow z/\sqrt{2}$ does not make sense. The factor of $e^{-z^{2}%
/4}$ has the effect of converting from the fully Gaussian measure in \cite{B}
to the measure $\nu_{\hbar},$ which is Gaussian in the imaginary directions
but constant in the real directions. (See \cite[Appendix]{H1}.)

The $C_{\hbar}$ form of the Segal--Bargmann transform has the advantage of
making explicit the symmetries of position-space. The measure $dx$ on
$\mathbb{R}^{d}$ and the measure $\nu_{\hbar}$ on $\mathbb{C}^{d}$ are both
invariant under rotations and translations of $x$-space, and the transform
commutes with rotations and translations of $x$-space. Since a gauge
transformation is just a combination of a rotation and a translation, this
property of $C_{\hbar}$ will be useful.

On the other hand, as it stands this form of the Segal--Bargmann transform
does not permit taking the infinite-dimensional limit, as we must do if we
want to quantize $\mathcal{A},$ since neither $dx$ nor $\nu_{\hbar}$ makes
sense when the dimension tends to infinity. Fortunately, it is not too hard to
fix this problem by adding a little bit of Gaussian-ness to our measures in
the $x$-directions. It turns out that if we do this correctly, then we can
keep the same formula for the Segal--Bargmann transform while making a small
change in the measures, and still have a unitary map. (See \cite[Sect. 3]%
{DH1}, \cite{H5}, or \cite{Sen}.)

\begin{theorem}
\label{sb2}For all $s>$ $\hbar/2$, let $P_{s}$ denote the measure on
$\mathbb{R}^{d}$ given by
\[
dP_{s}\left(  x\right)  =\left(  2\pi s\right)  ^{-d/2}e^{-x^{2}/2s}dx
\]
and let $M_{s,\hbar}$ denote the measure on $\mathbb{C}^{d}$ given by
\[
dM_{s,\hbar}\left(  x+ip\right)  =\left(  \pi\hbar\right)  ^{-d/2}\left(  \pi
r\right)  ^{-d/2}e^{-x^{2}/r}e^{-p^{2}/\hbar},
\]
where $r=2s-$ $\hbar.$ Then the map $S_{s,\hbar}:L^{2}(\mathbb{R}^{d}%
,P_{s})\rightarrow\mathcal{H}(\mathbb{C}^{d})$ given by
\[
S_{s,\hbar}f=\text{ analytic continuation of }e^{\hbar\Delta/2}f
\]
is a unitary map of $L^{2}(\mathbb{R}^{d},P_{s})$ onto $\mathcal{H}%
L^{2}(\mathbb{C}^{d},M_{s,\hbar}).$
\end{theorem}

If we multiply the measures on both sides by $\left(  2\pi s\right)  ^{d/2}$
and then let $s\rightarrow\infty$ we recover the $C_{\hbar}$ version of the
transform. On the other hand, for any finite value of $s$ it is possible to
let $d\rightarrow\infty$ to get a transform that is applicable to our
gauge-theory example. So we consider $L^{2}\left(  \mathcal{A},P_{s}\right)
,$ where $P_{s}$ is the Gaussian measure on $\mathcal{A}$ described in Section
3, which is just the infinite-dimensional limit of the measures $P_{s}$ on
$\mathbb{R}^{d}.$ We consider also the Gaussian measure $M_{s,\hbar}$ on
$\mathcal{A}_{\mathbb{C}}$ that is the infinite-dimensional limit of the
corresponding measures on $\mathbb{C}^{d}.$

We then work with \textbf{cylinder functions} in $L^{2}\left(  \mathcal{A}%
,P_{s}\right)  ,$ that is, functions that depend on only finitely many of the
infinitely many variables in $\mathcal{A}.$ (See \cite[Defn. 4.2]{DH1}.) On
such functions the Segal--Bargmann transform makes sense, since on such
functions $\Delta_{\mathcal{A}}$ reduces to the Laplacian for some
finite-dimensional space. It then follows from Theorem \ref{sb2} that the
Segal--Bargmann transform $S_{s,\hbar}$ is an isometric map of the space of
cylinder functions in $L^{2}\left(  \mathcal{A},P_{s}\right)  $ into
$\mathcal{H}L^{2}\left(  \mathcal{A}_{\mathbb{C}},M_{s,\hbar}\right)  .$ This
transform extends by continuity to a unitary map of $L^{2}\left(
\mathcal{A},P_{s}\right)  $ onto $\mathcal{H}L^{2}\left(  \mathcal{A}%
_{\mathbb{C}},M_{s,\hbar}\right)  .$ Recall that $\Delta_{\mathcal{A}}$ by
itself is a non-closable operator as a map of $L^{2}\left(  \mathcal{A}%
,P_{s}\right)  $ to itself. Considering $e^{\hbar\Delta_{\mathcal{A}}/2}$ as a
map from $L^{2}\left(  \mathcal{A},P_{s}\right)  $ to itself will not help
matters. But by considering $e^{\hbar\Delta_{\mathcal{A}}/2}$ followed by
analytic continuation, as a map from $L^{2}\left(  \mathcal{A},P_{s}\right)  $
to $\mathcal{H}L^{2}\left(  \mathcal{A}_{\mathbb{C}},M_{s,\hbar}\right)  ,$ we
get a map which is not only closable but continuous (even isometric). It then
makes perfect sense to apply this operator (the Segal--Bargmann transform) to
functions of the holonomy.

The following theorem summarizes the above discussion. (See Theorem 4.3 of
\cite{DH1}.)

\begin{theorem}
For all $s>$ $\hbar/2$ the map $S_{s,\hbar}$ given by
\[
S_{s,\hbar}f=\text{ analytic continuation of }e^{\hbar\Delta_{\mathcal{A}}%
/2}f
\]
makes sense and is isometric on cylinder functions, and extends by continuity
to a unitary map of $L^{2}\left(  \mathcal{A},P_{s}\right)  $ onto
$\mathcal{H}L^{2}\left(  \mathcal{A}_{\mathbb{C}},M_{s,\hbar}\right)  .$
\end{theorem}

We are now ready to state the main result (Theorem 5.2) of \cite{DH1}.

\begin{theorem}
\label{sb4}Suppose $f\in L^{2}(\mathcal{A},P_{s})$ is of the form
\[
f\left(  A\right)  =\phi\left(  h\left(  A\right)  \right)
\]
where $\phi$ is a function on $K.$ Then there exists a unique holomorphic
function $\Phi$ on $K_{\mathbb{C}}$ such that
\[
S_{s,\hbar}f\left(  C\right)  =\Phi\left(  h_{\mathbb{C}}\left(  C\right)
\right)  .
\]
The function $\Phi$ is given by
\[
\Phi=\text{analytic continuation }e^{\hbar\Delta_{K}/2}\phi.
\]
\end{theorem}

Note that in light of the definition of $S_{s,\hbar}$, this result says that
on the gauge-invariant subspace, $e^{\hbar\Delta_{\mathcal{A}}/2}$ (followed
by analytic continuation) reduces to $e^{\hbar\Delta_{K}/2}$ (followed by
analytic continuation). Thus Theorem \ref{sb4} is a formally equivalent to the
principle (\ref{main.idea}) with which we started. The $s=\hbar$ case of this
result is essentially due to Gross and Malliavin \cite{GM}. See also
\cite[Sect. 2.5]{HS} for more on the $s=\hbar$ case.

Now, the gauge-invariant subspace $L^{2}\left(  \mathcal{A},P_{s}\right)
^{\mathcal{G}_{0}}$ consists of functions of the form $f\left(  A\right)
=\phi\left(  h\left(  A\right)  \right)  ,$ with $\phi$ a function on $K.$ It
may be shown that
\[
\int_{\mathcal{A}}\left|  \phi\left(  h\left(  A\right)  \right)  \right|
^{2}\,dP_{s}\left(  A\right)  =\int_{K}\left|  \phi\left(  x\right)  \right|
^{2}\rho_{s}\left(  x\right)  \,dx,
\]
where $\rho_{s}$ is the heat kernel at the identity on $K$ at time $s.$
Similarly,
\[
\int_{\mathcal{A}_{\mathbb{C}}}\left|  \Phi\left(  h_{\mathbb{C}}\left(
Z\right)  \right)  \right|  ^{2}\,dM_{s,\hbar}\left(  Z\right)  =\int
_{K_{\mathbb{C}}}\left|  \Phi\left(  g\right)  \right|  ^{2}\mu_{s,\hbar
}\left(  g\right)  \,dg,
\]
where $\mu_{s,\hbar}$ is a suitable heat kernel on $K_{\mathbb{C}}$ and $dg$
is Haar measure on $K_{\mathbb{C}}.$ So the gauge-invariant subspace on the
real side is identifiable with $L^{2}\left(  K,\rho_{s}\left(  x\right)
\,dx\right)  $ and on the complex side with $\mathcal{H}L^{2}\left(
K_{\mathbb{C}},\mu_{s,\hbar}\left(  g\right)  \,dg\right)  .$ So we have the
following commutative diagram in which all maps are unitary.
\begin{equation}%
\begin{array}
[c]{ccc}%
L^{2}\left(  \mathcal{A},P_{s}\right)  ^{\mathcal{G}_{0}} & \underrightarrow
{e^{\hbar\Delta_{\mathcal{A}}/2}} & \mathcal{H}L^{2}(\mathcal{A}_{\mathbb{C}%
},M_{s,\hbar})^{\mathcal{G}_{0}}\\
\updownarrow &  & \updownarrow\\
L^{2}\left(  K,\rho_{s}\left(  x\right)  dx\right)  & \underrightarrow
{e^{\hbar\Delta_{K}/2}} & \mathcal{H}L^{2}(K_{\mathbb{C}},\mu_{s,\hbar}\left(
g\right)  dg)
\end{array}
\label{commute}%
\end{equation}
The horizontal maps contain an implicit analytic continuation.

This result embodies a rigorous version of the principle (\ref{main.idea}) and
also shows that a form of the Segal--Bargmann transform for $\mathcal{A}$ can
descend to a Segal--Bargmann transform for $\mathcal{A}/\mathcal{G}_{0}=K.$
But so far we still have the regularization parameter $s,$ which we are
supposed to remove by letting it tend to infinity. On the full space
$L^{2}\left(  \mathcal{A},P_{s}\right)  $ or $\mathcal{H}L^{2}(\mathcal{A}%
_{\mathbb{C}},M_{s,\hbar})$ the limit $s\rightarrow\infty$ does not exist;
this was the point of putting in the $s$ in the first place. But on the
gauge-invariant subspaces, identified with functions on $K$ or $K_{\mathbb{C}%
},$ the limit does exist. As $s\rightarrow\infty,$ the heat kernel measure
$\rho_{s}$ on $K$ converges to normalized Haar measure on $K.$ This confirms
our earlier conjecture that the fictitious Lebesgue measure on $\mathcal{A}$
(formally the $s\rightarrow\infty$ limit of $P_{s}$) pushes forward to the
Haar measure on $K.$ Meanwhile, the measure $\mu_{s,\hbar}$ converges as
$s\rightarrow\infty$ to a certain measure I call $\nu_{\hbar},$ which
coincides with the ``$K$-averaged heat kernel measure'' of \cite{H1}. So
taking the limit in the bottom line of (\ref{commute}) gives a unitary map
\begin{equation}%
\begin{array}
[c]{ccc}%
L^{2}\left(  K,dx\right)  & \underrightarrow{e^{\hbar\Delta_{K}/2}} &
\mathcal{H}L^{2}(K_{\mathbb{C}},\nu_{\hbar})
\end{array}
\label{commute2}%
\end{equation}
This supports the expected conclusion that our reduced quantum Hilbert space
is $L^{2}\left(  K,dx\right)  $ and that the quantum Hamiltonian is $\left(
-\hbar^{2}/2\right)  \Delta_{K}.$ It further shows that the generalized
Segal--Bargmann transform for $K,$ as given in (\ref{commute2}), arises
naturally from the ordinary Segal--Bargmann transform for the space of
connections, upon restriction to the gauge-invariant subspace. The transform
in (\ref{commute2}) is precisely the $K$-invariant form $C_{\hbar}$ of the
transform, as previously constructed in \cite{H1} from a purely
finite-dimensional point of view.

\section{Coherent states: from $\mathcal{A}_{\mathbb{C}}$ to $K_{\mathbb{C}}$}

Let us now reformulate the results of the last section in terms of coherent
states. Described in this way, our results are in the spirit of the proposal
of J. Klauder \cite{Kl1,Kl2} on how to quantize systems with constraints. (See
also \cite{GK}.)

Klauder and B.-S. Skagerstam \cite{KS} think of coherent states as a
collection of states $\psi_{\alpha}$ in some Hilbert space $H$, labeled by
points $\alpha$ in some parameter space $X.$ They assume that there is a
``resolution of the identity''
\begin{equation}
I=\int_{X}\left|  \psi_{\alpha}\right\rangle \left\langle \psi_{\alpha
}\right|  \,d\nu\left(  \alpha\right) \label{res.id}%
\end{equation}
for some measure $\nu$ on $X.$ One may then define a ``coherent state
transform,'' that is, a linear map $C:H\rightarrow L^{2}\left(  X,\nu\right)
$ given by taking the inner product of a vector in $H$ with each of the
coherent states:
\[
C\left(  v\right)  \left(  \alpha\right)  =\left\langle \psi_{\alpha}\left|
v\right.  \right\rangle .
\]

The resolution of the identity implies that
\begin{align*}
\int_{X}\left|  \left\langle \psi_{\alpha}\left|  v\right.  \right\rangle
\right|  ^{2}\,d\nu\left(  \alpha\right)   & =\int_{X}\left\langle v\left|
\psi_{\alpha}\right.  \right\rangle \left\langle \psi_{\alpha}\left|
v\right.  \right\rangle \,d\nu\left(  \alpha\right) \\
& =\left\langle v\right|  \int_{X}\left|  \psi_{\alpha}\right\rangle
\left\langle \psi_{\alpha}\right|  \,d\nu\left(  \alpha\right)  \left|
v\right\rangle \\
& =\left\langle v\left|  v\right.  \right\rangle .
\end{align*}
Thus the resolution of the identity (\ref{res.id}) is equivalent to the
statement that $C$ is an isometric linear map. Note that $C$ is only isometric
but \textit{not} unitary. In all the interesting cases the image of $C$ is a
proper subspace of $L^{2}\left(  X,\nu\right)  ,$ which may be characterized
by a certain reproducing kernel condition. Although the resolution of the
identity looks on the surface like an orthonormal basis expansion, it is in
fact quite different. The coherent states are typically non-orthogonal and
``overcomplete.'' The overcompleteness is reflected in the fact that $C$ does
not map onto $L^{2}\left(  X,\nu\right)  .$

As an example, consider the finite-dimensional Segal--Bargmann transform, in
my normalization. The coherent states are then the states $\psi_{z}\in
L^{2}(\mathbb{R}^{d},dx)$ given by
\[
\psi_{z}\left(  x\right)  =\left(  2\pi\hbar\right)  ^{-n/2}e^{-\left(
\bar{z}-x\right)  ^{2}/2\hbar},\quad z\in\mathbb{C}^{n}.
\]
This means that the coherent state transform is given by
\[
\left(  C_{\hbar}f\right)  \left(  z\right)  =\left\langle \psi_{z}\left|
f\right.  \right\rangle _{L^{2}(\mathbb{R}^{d},dx)}=\int_{\mathbb{R}^{d}%
}\left(  2\pi\hbar\right)  ^{-n/2}e^{-\left(  z-x\right)  ^{2}/2\hbar}f\left(
x\right)  \,dx,
\]
as above. In this case the parameter space $X$ is $\mathbb{C}^{d}$ and the
measure on $X$ is the measure $\nu_{\hbar}$ of the last section. The
overcompleteness of the coherent states means here that the image of
$C_{\hbar}$ is not all of $L^{2}(\mathbb{C}^{d},\nu_{\hbar}),$ but only the
holomorphic subspace.

Next consider what happens to a set of coherent states under reduction.
Suppose we have a set of coherent states in a Hilbert space $H,$ satisfying a
resolution of the identity (\ref{res.id}). Then suppose that $V$ is a closed
subspace of $H$ and that $P$ is the orthogonal projection onto $V.$ Since
$P^{2}=P^{\ast}=P,$ (\ref{res.id}) gives
\[
P=PIP=\int_{X}\left|  P\psi_{\alpha}\right\rangle \left\langle P\psi_{\alpha
}\right|  \,d\nu\left(  \alpha\right)  .
\]
Thus by projecting each coherent state into $V$ we get a resolution of the
identity (and hence a coherent state transform) for the subspace $V.$ The
``reduced coherent states'' are the projections $P\psi_{\alpha}$ of the
original coherent states into the subspace $V.$

Note that at the moment the parameter space for the coherent states, and the
measure on it, are unchanged by the projection. However, it may happen that
certain sets of distinct coherent states become the same after the projection
is applied. In that case we may reduce (or ``collapse'') the parameter space
$X$ by identifying any two parameters $\alpha$ and $\beta$ for which
$P\psi_{\alpha}=P\psi_{\beta}.$ The measure $\nu$ then pushes forward to a
measure $\tilde{\nu}$ on the reduced parameter space $\tilde{X}. $

This is what happens in our Yang-Mills case. Initially we have coherent states
$\psi_{Z}\in L^{2}\left(  \mathcal{A},P_{s}\right)  $ labeled by points $Z$ in
the phase space $\mathcal{A}_{\mathbb{C}}.$ These states satisfy
\begin{equation}
S_{s,\hbar}f\left(  Z\right)  =\left\langle \left.  \psi_{Z}\right|
f\right\rangle _{L^{2}\left(  \mathcal{A},P_{s}\right)  },\label{state.def}%
\end{equation}
I suppress the dependence of $\psi_{Z}$ on $s$ and $\hbar.$ We want to project
the $\psi_{Z}$'s onto the gauge-invariant subspace, that is, onto the space of
functions of the form $\phi\left(  h\left(  A\right)  \right)  .$ The
projection amounts to the same thing as restricting attention in
(\ref{state.def}) to $f$'s of the form $f\left(  A\right)  =\phi\left(
h\left(  A\right)  \right)  .$ For such $f$'s, Theorem \ref{sb4} tells us
that
\begin{align*}
\left\langle \left.  \psi_{Z}\right|  f\right\rangle  & =\left[  S_{s,\hbar
}\left(  \phi\circ h\right)  \right]  \left(  Z\right) \\
& =\Phi\left(  h_{\mathbb{C}}\left(  Z\right)  \right)  ,
\end{align*}
where $\Phi$ is the analytic continuation to $K_{\mathbb{C}}$ of
$e^{\hbar\Delta_{K}/2}\phi.$ We see then that for $f$ in the invariant
subspace, the right side of (\ref{state.def}) depends only on the holonomy of
$Z.$ This says that if we have two different coherent states $\psi_{Z}$ and
$\psi_{W}$ such that $h_{\mathbb{C}}\left(  Z\right)  =h_{\mathbb{C}}\left(
W\right)  ,$ then upon projection into the gauge-invariant subspace, they will
become equal. Thus the parameter space for the coherent states ``collapses''
from $\mathcal{A}_{\mathbb{C}}$ to $\mathcal{A}_{\mathbb{C}}/\mathcal{G}%
_{0,\mathbb{C}}=K_{\mathbb{C}}.$

If we identify the gauge-invariant subspace with $L^{2}\left(  K,\rho
_{s}\right)  $ as in the previous section, then the reduced coherent states
are the vectors $\tilde{\psi}_{g}\in L^{2}\left(  K,\rho_{s}\right)  ,$ with
$g\in K_{\mathbb{C}},$ given by
\[
\tilde{\psi}_{g}\left(  x\right)  =\frac{\overline{\rho_{\hbar}\left(
gx^{-1}\right)  }}{\rho_{s}\left(  x\right)  },\quad g\in K_{\mathbb{C}},
\]
so that, as required, we have
\begin{align*}
\left\langle \left.  \tilde{\psi}_{g}\right|  \phi\right\rangle _{L^{2}\left(
K,\rho_{s}\right)  }  & =\int_{K}\frac{\rho_{\hbar}\left(  gx^{-1}\right)
}{\rho_{s}\left(  x\right)  }\phi\left(  x\right)  \,\rho_{s}\left(  x\right)
\,dx\\
& =\Phi\left(  g\right)  .
\end{align*}
Here $\rho_{\hbar}\left(  gx^{-1}\right)  $ refers to the analytic
continuation of the heat kernel from $K$ to $K_{\mathbb{C}},$ and for $g\in
K,$ the convolution $\int_{K}\rho_{\hbar}\left(  gx^{-1}\right)  \phi\left(
x\right)  \,dx$ is nothing but $(e^{\hbar\Delta_{K}/2}\phi)\left(  g\right)  .$

The $\tilde{\psi}_{g}$'s satisfy a resolution of the identity with respect to
the measure $\mu_{s,\hbar}$ on $K_{\mathbb{C}}.$ This measure is the one which
is naturally induced from the measure $M_{s,\hbar}$ on $\mathcal{A}%
_{\mathbb{C}},$ upon reduction from $\mathcal{A}_{\mathbb{C}}$ to
$K_{\mathbb{C}}.$ That is, $\mu_{s,\hbar}$ is the ``push-forward'' of
$M_{s,\hbar}$ from $\mathcal{A}_{\mathbb{C}}$ to $K_{\mathbb{C}},$ under the
map $h_{\mathbb{C}}.$ Now, as $s\rightarrow\infty,$ $\rho_{s}\left(  x\right)
$ converges to the constant function $\mathbf{1}$. Thus we obtain in the limit
coherent states $\chi_{g}\in L^{2}\left(  K,dx\right)  $ given by
\begin{equation}
\chi_{g}\left(  x\right)  :=\lim_{s\rightarrow\infty}\frac{\overline
{\rho_{\hbar}\left(  gx^{-1}\right)  }}{\rho_{s}\left(  x\right)  }%
=\overline{\rho_{\hbar}\left(  gx^{-1}\right)  },\quad g\in K_{\mathbb{C}%
}.\label{group.coh}%
\end{equation}
These satisfy the following resolution of the identity:
\[
I=\int_{K_{\mathbb{C}}}\left|  \chi_{g}\right\rangle \left\langle \chi
_{g}\right|  \,d\nu_{\hbar}\left(  g\right)  ,
\]
where $\nu_{\hbar}=\lim_{s\rightarrow\infty}\mu_{s,\hbar}.$ The measure
$\nu_{\hbar}$ coincides with the ``$K$-averaged heat kernel measure'' of
\cite{H1}.

Although we are ``supposed to'' let $s\rightarrow\infty,$ we get a
well-defined family of coherent states for any $s>$ $\hbar/2.$ The case $s=$
$\hbar$, as well as the limiting case $s\rightarrow\infty,$ had previously
been described in \cite{H1}. For other values of $s$ we get something new,
which I investigate from a finite-dimensional point of view in \cite{H5}.

Let me compare the above results to those in the paper of Wren \cite{W}, which
motivated Driver and me to develop our paper \cite{DH1}. Wren uses the
``Rieffel induction'' method proposed by Landsman \cite{L1}, applied to this
same problem of Yang-Mills theory on a spacetime cylinder. The commutative
case was considered previously by Landsman and Wren in \cite{LW}. Wren uses a
fixed Gaussian measure and a ``unitarized'' action of the gauge group. In this
approach there is no gauge-invariant subspace (see \cite{DH2}) and so an
integration over the gauge group is used to define a reduced Hilbert space,
which substitutes for the gauge-invariant subspace. Wren shows that the
reduced Hilbert space can be identified with $L^{2}\left(  K,dx\right)  $ and
further shows that under the reduction map the ordinary coherent states map
precisely to the coherent states $\chi_{g}$ in (\ref{group.coh}). So the
appearance of these coherent states in \cite{DH1} was expected on the basis of
Wren's results.

The paper \cite{DH1} set out to understand better two issues raised by
\cite{W}. First, because in \cite{W} there is no true gauge-invariant subspace
to project onto, the resolution of the identity for the classical coherent
states does not survive the reduction. That is, Rieffel induction does not
tell you what measure to use on $K_{\mathbb{C}}$ in order to get a resolution
of the identity. Of course, the relevant measure had already been described in
\cite{H1}, but it would be nice not to have to know this ahead of time. By
contrast, in our approach the measure $\nu_{\hbar}$ arises naturally by
pushing forward the Gaussian measure $M_{s,\hbar}$ to $K_{\mathbb{C}}$ and
then letting $s$ tend to infinity. Second, the calculation in \cite{W}
concerning the reduction of the Hamiltonian is non-rigorous, mainly because
the \textit{unconstrained} Hamiltonian is not well-defined. Driver and I used
the Segal--Bargmann transform in order to get some form of the Hamiltonian
$\Delta_{\mathcal{A}}$ to make rigorous sense.

Finally, let me mention that the generalized coherent states on $K$ are do not
fall into the framework of Perelomov \cite{P}, because there does not seem to
be in the compact group case anything analogous to the irreducible unitary
representation of the Heisenberg group on $L^{2}(\mathbb{R}^{d})$.

\section{Identification of $T^{\ast}\!(K)$ with $K_{\mathbb{C}}$}

I am thinking of $K$ as the configuration space for the reduced classical
Yang-Mills theory, and of $K_{\mathbb{C}}$ as the corresponding phase space.
For this to be sensible, there should be an identification of $K_{\mathbb{C}}$
with the standard phase space over $K,$ namely the cotangent bundle $T^{\ast
}\!(K).$ In this section I will explain how such an identification comes out
of the reduction process. The resulting identification coincides with the one
constructed in \cite{H3,H4} from an intrinsic point of view.

Let us see what comes out of the reduction process. From the symplectic point
of view we have the Marsden--Weinstein symplectic quotient $J^{-1}\left(
0\right)  /\mathcal{G}_{0}.$ Since the action of $\mathcal{G}_{0} $ on
$\mathcal{A}_{\mathbb{C}}=T^{\ast}\!(\mathcal{A})$ arises from an action of
$\mathcal{G}_{0}$ on the configuration space $\mathcal{A},$ general principles
tell us that $J^{-1}\left(  0\right)  /\mathcal{G}_{0}$ coincides with
$T^{\ast}\!\left(  \mathcal{A}/\mathcal{G}_{0}\right)  =T^{\ast}\!(K).$ On the
other hand, from the complex point of view we may analytically continue the
action of $\mathcal{G}_{0}$ on $\mathcal{A}_{\mathbb{C}}$ to get an action of
$\mathcal{G}_{0,\mathbb{C}}$ on $\mathcal{A}_{\mathbb{C}}.$ Dividing out by
this action gives $\mathcal{A}_{\mathbb{C}}/\mathcal{G}_{0,\mathbb{C}%
}=K_{\mathbb{C}}.$ But in this case there is a natural identification of
$J^{-1}\left(  0\right)  /\mathcal{G}_{0}$ with $\mathcal{A}_{\mathbb{C}%
}/\mathcal{G}_{0,\mathbb{C}}$: each orbit of $\mathcal{G}_{0,\mathbb{C}}$
intersects $J^{-1}\left(  0\right)  $ in precisely one $\mathcal{G}_{0}%
$-orbit. This may be seen from \cite[Sect. IV.3.6]{L2}.

This result is not a coincidence. In general, given a K\"{a}hler manifold $M$
(in our example $\mathcal{A}_{\mathbb{C}}$) and an action of a group $G$ that
preserves both the complex and symplectic structure of $M,$ we may
analytically continue to get an action of $G_{\mathbb{C}}$ on $M,$ an action
which preserves the complex but not the symplectic structure of $M.$ Then if
$J^{-1}\left(  0\right)  $ is the moment mapping for the action of $G,$ one
expects that
\begin{equation}
J^{-1}\left(  0\right)  /G=M/G_{\mathbb{C}}.\label{mod.gc}%
\end{equation}
This would mean that for each orbit $O$ of $G_{\mathbb{C}}$ in $M$ the
intersection of $O$ with $J^{-1}\left(  0\right)  $ is precisely a single
$G$-orbit. Now, (\ref{mod.gc}) is not actually true in general, but only with
various provisos and qualifications. (See \cite{Ki,MFK} and the notes in
Section 9.) Still, this is an important idea and in our case it works out exactly.

Putting everything together we have the following identifications.
\[%
\begin{array}
[c]{ccccc}%
\mathcal{A}_{\mathbb{C}}/\mathcal{G}_{0,\mathbb{C}} & = & J^{-1}\left(
0\right)  /\mathcal{G}_{0} & = & T^{\ast}\!(\mathcal{A}/\mathcal{G}_{0})\\
\updownarrow &  &  &  & \updownarrow\\
K_{\mathbb{C}} &  &  &  & T^{\ast}\!(K)
\end{array}
\]
If one does the calculations, one obtains the following explicit
identification of $T^{\ast}\!(K)$ with $K_{\mathbb{C}}.$ (See Proposition
3.6.8 of Landsman \cite[Sect. IV.3.6]{L2}. Landsman uses slightly different
conventions.) First, use left-translation to trivialize the cotangent bundle,
so that $T^{\ast}\!(K)\cong K\times\frak{k}^{\ast}.$ Then use the inner
product on $\frak{k}$ to identify $K\times\frak{k}^{\ast}$ with $K\times
\frak{k}.$ Finally map from $K\times\frak{k}$ to $K_{\mathbb{C}}$ by the map
\begin{equation}
\Phi\left(  x,Y\right)  =xe^{iY},\quad x\in K,\,Y\in\frak{k}.\label{polar}%
\end{equation}
The map $\Phi$ is a diffeomorphism of $K\times\frak{k}$ onto $K_{\mathbb{C}},$
and $\Phi^{-1}$ is called the polar decomposition of $K_{\mathbb{C}}.$

Of course, one could simply write down (\ref{polar}) directly at the
finite-dimensional level, and indeed this is what I do in \cite[Sect. 3]{H3}.
However, it is interesting that this same identification comes out naturally
from the reduction process (along with the Segal--Bargmann transform).

To illustrate the identification of $K_{\mathbb{C}}$ with $T^{\ast}\!(K),$
consider the case $K=SU(n),$ in which case $K_{\mathbb{C}}=SL(n;\mathbb{C}).$
Then for any $g$ in $SL\left(  n;\mathbb{C}\right)  $ we may use the standard
polar decomposition for matrices to write
\[
g=xp
\]
with $x$ unitary and $p$ positive. Since $\det g=1$ it follows that $\det x=1
$ and $\det p=1$ (since $\det x$ has absolute value one and $\det p$ is real
and positive). In particular, $x\in SU(n).$ Then $p$ has a unique self-adjoint
logarithm $\xi$, which has trace zero. Letting $Y=\xi/i$ we have
\[
g=xe^{iY},
\]
where $Y$ is skew and has trace zero, i.e. $Y$ is in $su\left(  n\right)  .$
Thus we see that $SL(n;\mathbb{C})$ decomposes as $SU\left(  n\right)  \times
su\left(  n\right)  \cong T^{\ast}(SU(n))$ as in (\ref{polar}).

Now in \cite[Sect. 3]{H3} (see also \cite{H4}) I argued from an intrinsic,
finite-dimensional point of view that the above identification of $T^{\ast
}\!(K)$ with $K_{\mathbb{C}}$ was natural. The argument was based on the
notion of ``adapted complex structures'' \cite{GStenz1,GStenz2,LS,Sz}. There
is a good reason that the reduction argument gives the same identification as
the adapted complex structures do. Suppose $X$ is a finite-dimensional compact
Riemannian manifold such that $T^{\ast}\!\left(  X\right)  $ has a global
adapted complex structure, and suppose $G$ is a compact Lie group which acts
freely and isometrically on $X.$ Then a result of R. Aguilar \cite{A} says
that $T^{\ast}\!(X/G)$ has a global adapted complex structure and that this
complex structure coincides with the one inherited from $T^{\ast}\!(X)$ by
means of reduction. We have the same sort of situation here, with
$X=\mathcal{A}$ and $G=\mathcal{G}_{0}.$ Of course, $\mathcal{G}_{0}$ is not
compact and $\mathcal{A}$ is neither compact nor finite-dimensional, but
nevertheless what happens is reasonable in light of Aguilar's result.

\section{Reduction of the Laplacian}

Why \textit{should} $\Delta_{\mathcal{A}}$ correspond to $\Delta_{K}$ on
gauge-invariant functions? Let us strip away the infinite-dimensional
technicalities and consider the analogous question in finitely many
dimensions. Suppose $X$ is a finite-dimensional connected Riemannian manifold
and suppose $G$ is a Lie group that acts by isometries on $X.$ For simplicity
I will assume that $G$ is compact and that $G$ acts freely on $X.$ Then $X/G$
is again a manifold, which has a unique Riemannian metric such that the
quotient map $q:X\rightarrow X/G$ is a Riemannian submersion. This means that
at each point $x\in X$, the differential of $q$ is an isometry when restricted
to the orthogonal complement of the tangent space to the $G$-orbit through $x.$

Given this ``submersion'' metric on $X/G$ we may consider the Laplace-Beltrami
operator $\Delta_{X/G}$. For a smooth function $f$ on $X/G$ we may ask whether
$\left(  \Delta_{X/G}f\right)  \circ q$ coincides with $\Delta_{X}\left(
f\circ q\right)  .$ This amounts to asking whether $\Delta_{X}$ and
$\Delta_{X/G}$ agree on the $G$-invariant subspace of $C^{\infty}\left(
X\right)  .$ Since $\Delta_{X}$ commutes with isometries, it will at least
preserve the $G$-invariant subspace.

The answer in general is no, $\Delta_{X}$ and $\Delta_{X/G}$ do \textit{not}
agree on $C^{\infty}\left(  X\right)  ^{G}.$ For example, consider $SO\left(
2\right)  $ acting on $\mathbb{R}^{2}\setminus\left\{  0\right\}  $ by
rotations. The quotient manifold is diffeomorphic to $\left(  0,\infty\right)
,$ with the point $r\in\left(  0,\infty\right)  $ corresponding to the orbit
$x^{2}+y^{2}=r^{2}$ in $\mathbb{R}^{2}\setminus\left\{  0\right\}  .$ The
submersion metric on $\left(  0,\infty\right)  $ is the usual metric on
$\left(  0,\infty\right)  $ as a subset of $\mathbb{R}.$ So the
Laplace-Beltrami operator on $\left(  0,\infty\right)  $ is just $d^{2}%
/dr^{2}.$ On the other hand, the formula for the two-dimensional Laplacian on
radial functions is
\begin{equation}
\left(  \frac{\partial^{2}}{\partial x^{2}}+\frac{\partial^{2}}{\partial
y^{2}}\right)  f\left(  \sqrt{x^{2}+y^{2}}\right)  =\left.  \left[
\frac{d^{2}f\left(  r\right)  }{dr^{2}}+\frac{1}{r}\frac{df\left(  r\right)
}{dr}\right]  \right|  _{r=\sqrt{x^{2}+y^{2}}},\label{laps}%
\end{equation}
which differs from $d^{2}f/dr^{2}$ by a first-order term. The source of the
trouble is the discrepancy between the intrinsic volume measure $dr$ on
$\left(  0,\infty\right)  $ and the push-forward of the volume measure from
$\mathbb{R}^{2}\setminus\left\{  0\right\}  ,$ which is $2\pi r\,dr.$

In general, each $G$-orbit in $X$ inherits a natural Riemannian metric from
$X$, and we may compute the total volume of this orbit with respect to the
associated Riemannian volume measure. The expression $\mathrm{Vol}\left(
G\cdot x\right)  $ is a function on $X/G,$ and it measures the discrepancy
between the intrinsic volume measure on $X/G$ and the push-forward of the
volume measure on $X.$ The two Laplacians on $C^{\infty}\left(  X\right)  ^{G}
$ will be related by the formula
\begin{equation}
\Delta_{X}=\Delta_{X/G}+\nabla\left(  \log\mathrm{Vol}\left(  G\cdot x\right)
\right)  \cdot\nabla.\label{laps2}%
\end{equation}
(The gradient may be thought of as that for $X/G$, although this coincides in
a natural sense with that for $X,$ on $G$-invariant functions.) Formula
(\ref{laps}) is a special case of (\ref{laps2}) with volume factor $2\pi r.$
We have arrived at the following conclusion.

\begin{quote}
The Laplacians $\Delta_{X}$ and $\Delta_{X/G}$ agree on $G$-invariant
functions if and only if all the $G$-orbits have the same volume.
\end{quote}

Let us return, then, to the case of $\mathcal{A}/\mathcal{G}_{0}.$ By
considering the appendix of \cite{DH1} it is easily seen that the metric on
$K$ that makes the map $h:\mathcal{A}\rightarrow K$ a Riemannian submersion is
simply the bi-invariant metric on $K$ induced by the inner product on
$\frak{k}.$ (We use on $\mathcal{A}$ the metric coming from the $L^{2}$ norm
as in Section 2.) So in light of (\ref{laps2}) the statement that
$\Delta_{\mathcal{A}}$ and $\Delta_{K}$ agree on the $\mathcal{G}_{0}%
$-invariant subspace is formally equivalent to the statement that all the
$\mathcal{G}_{0}$-orbits have the same volume. In this case (for connections
on a circle) it can be shown that there exist isometries of $\mathcal{A}$ that
map any $\mathcal{G}_{0}$-orbit to any other. Thus formally all the
$\mathcal{G}_{0}$-orbits should have the same volume.

The relevant isometries come from extending the gauge action (\ref{action}) of
$\mathcal{G}_{0}$ (the loop group over $K$) on $\mathcal{A}$ to an action of
the pathgroup over $K,$ given by the same formula. This action of the
pathgroup on $\mathcal{A}$ is isometric. If $g_{\tau}$ is a path in $K$ with
$g_{0}=e$ but with $g_{1}$ arbitrary, then $g$ changes holonomies according to
the formula $h\left(  g\cdot A\right)  =h\left(  A\right)  g_{1}^{-1}.$ Thus
the pathgroup permutes the $\mathcal{G}_{0}$-orbits (labeled by the holonomy),
and any $\mathcal{G}_{0}$-orbit can be mapped to any other by an element of
the pathgroup.

To look at the problem in another way, to see that $\Delta_{\mathcal{A}}$
matches up with $\Delta_{K}$ on gauge-invariant functions we need to show that
the (fictitious) volume measure $\mathcal{D}A$ on $\mathcal{A}$ pushes forward
to a constant multiple of the Haar measure on $K$. After all, the density of
the push-forward of $\mathcal{D}A$ with respect to the Haar measure should be
the volume factor, which we want to show is constant. If we accept the
Gaussian measures $P_{s}$ as an approximation to $\mathcal{D}A, $ we note that
these push forward to the measures $\rho_{s}\left(  x\right)  \,dx,$ which
indeed converge to Haar measure as $s$ tends to infinity.

It should be emphasized that these arguments apply only in our 1+1-dimensional
example. If $\mathcal{A}$ is the space of connections over some space manifold
$\mathcal{M}$ with dimension at least two, then $\Delta_{\mathcal{A}}$ and
$\Delta_{\mathcal{A}/\mathcal{G}_{0}}$ will \textit{not} coincide (even
formally) on gauge-invariant functions, as shown by K. Gaw\c{e}dzki \cite{Ga}.

\section{Does quantization commute with reduction?}

When quantizing a reduced Hamiltonian system such as Yang-Mills theory, one
may ask whether the quantization should be done before or after the reduction.
If we were very optimistic, we might hope that it doesn't matter, that one
gets the same answer either way. If this were so, we could say that
\textit{quantization commutes with reduction}. Of course the question of
whether quantization commutes with reduction may well depend on the system
being quantized and on how one interprets the question. I want to consider
this question from the point of view of geometric quantization \cite{H7,H8}
and I want specifically to compare the Segal--Bargmann space obtained by first
quantizing $\mathcal{A}_{\mathbb{C}}$ and then reducing by $\mathcal{G}_{0}$
to the one obtained by directly quantizing $K_{\mathbb{C}}.$

In geometric quantization \cite{Wo} one begins with a symplectic manifold
$\left(  M,\omega\right)  $ and constructs over $M$ a Hermitian complex line
bundle $L$ with connection, whose curvature form is $i\omega/2\pi\hbar.$ If
$M$ is a cotangent bundle then such a bundle exists and may be taken to be
topologically and Hermitianly trivial (though the connection is necessarily
non-trivial). The ``prequantum Hilbert space'' is then the space of sections
of $L$ which are square-integrable with respect to the symplectic volume
measure on $M.$ To obtain the ``quantum Hilbert space'' one picks a
``polarization'' and restricts to the space of square-integrable polarized
sections of $L.$ If $M$ is a K\"{a}hler manifold, i.e. it has a complex
structure which is compatible in a natural sense with $\omega,$ then there is
a natural K\"{a}hler polarization. In that case $L$ may be given the structure
of a holomorphic line bundle and the quantum Hilbert space becomes the space
of square-integrable holomorphic sections of $L.$

In the case $M=\mathbb{C}^{d}$ the resulting bundle is holomorphically
trivial. So by choosing a nowhere vanishing holomorphic section, the space of
holomorphic sections of $L$ may be identified with the space of holomorphic
functions on $\mathbb{C}^{d}.$ This nowhere vanishing section will not,
however, have constant norm. This means that the inner product on the space of
holomorphic functions will be an $L^{2}$ inner product with respect to a
measure which is Lebesgue measure times the norm-squared of the trivializing
section. Working this out we get simply the Segal--Bargmann space, with
different normalizations of the space coming from different choices of the
trivializing section. In summary: applying geometric quantization to
$\mathbb{C}^{d},$ using a K\"{a}hler polarization, yields the Segal--Bargmann space.

To apply geometric quantization to the infinite-dimensional space
$\mathcal{A}_{\mathbb{C}}$ we may try to quantize $\mathbb{C}^{d}$ and then
let $d$ tend to infinity. For this to make sense with my normalization, we
need to add the additional parameter $s.$ So we obtain the Segal--Bargmann
space $\mathcal{H}L^{2}(\mathcal{A}_{\mathbb{C}},M_{s,\hbar}).$ We then want
to reduce by the action of $\mathcal{G}_{0},$ which amounts to restricting to
the space of functions in $\mathcal{H}L^{2}(\mathcal{A}_{\mathbb{C}%
},M_{s,\hbar})$ which are $\mathcal{G}_{0}$-invariant, and thus by
analyticity, $\mathcal{G}_{0,\mathbb{C}}$-invariant. The resulting space is
identifiable with $\mathcal{H}L^{2}(K_{\mathbb{C}},\mu_{s,\hbar}).$ Finally,
letting $s$ tend to infinity we obtain $\mathcal{H}L^{2}\left(  K_{\mathbb{C}%
},\nu_{\hbar}\right)  .$ It is therefore reasonable to say that $\mathcal{H}%
L^{2}\left(  K_{\mathbb{C}},\nu_{\hbar}\right)  $ is the space obtained by
quantizing $\mathcal{A}_{\mathbb{C}} $ and then reducing by $\mathcal{G}_{0}.$

Alternatively, we may do the reduction first, obtaining the symplectic
manifold $T^{\ast}\!(K).$ This may be made into a K\"{a}hler manifold by
identifying $T^{\ast}\!(K)$ with $K_{\mathbb{C}}$ using the polar
decomposition, as described in Section 6 and in \cite[Section 3]{H3}. We may
then apply geometric quantization directly to $K_{\mathbb{C}}\cong T^{\ast
}\!(K).$ I do this calculation in \cite{H4} and find that geometric
quantization yields the space $\mathcal{H}L^{2}\left(  K_{\mathbb{C}}%
,\gamma_{\hbar}\right)  ,$ where $\gamma_{\hbar}$ and $\nu_{\hbar}$ are
related by the formula
\begin{equation}
d\nu_{\hbar}\left(  g\right)  =a_{\hbar}u\left(  g\right)  \,d\gamma_{\hbar
}\left(  g\right)  .\label{quant}%
\end{equation}
Here $a_{\hbar}$ is an irrelevant constant and $u$ is a function which is
non-constant except when $K$ is commutative. So it seems that quantizing
$K_{\mathbb{C}}$ directly does \textit{not} yield the same answer as
quantizing $\mathcal{A}_{\mathbb{C}}$ first and then reducing by
$\mathcal{G}_{0}.$

However, this is not the end of the story. One can quantize $K_{\mathbb{C}}$
taking into account the ``half-form correction'' (also known as the
``metaplectic correction''). This ``corrected'' quantization yields an extra
factor in the measure, a factor that coincides precisely with the factor
$u\left(  g\right)  $ in (\ref{quant})! (See \cite{H7,H8}.) On the other hand,
in the $\mathbb{C}^{n}$ case the half-form correction does not affect the
Hilbert space, so even with the half-form correction we would get
$\mathcal{H}L^{2}(\mathcal{A}_{\mathbb{C}},M_{s,\hbar})$ and then after
reduction $\mathcal{H}L^{2}\left(  K_{\mathbb{C}},\nu_{\hbar}\right)  .$ So
our conclusion is the following: In this example, if we use geometric
quantization with a K\"{a}hler polarization and the half-form correction,
quantization does in fact commute with reduction.

Let me conclude by mentioning a related setting in which one can ask whether
quantization commutes with reduction. In an influential paper \cite{GStern},
V. Guillemin and S. Sternberg consider the geometric quantization of a compact
K\"{a}hler manifold $M.$ They assume then that there is an action of a compact
group $G$ on $M$ that preserves the complex structure and the symplectic
structure of $M$ and they consider as well the Marsden--Weinstein quotient
$M^{G}:=J^{-1}\left(  0\right)  /G,$ where $J$ is the moment mapping for the
action of $G.$ Under certain conditions they show that there is a natural
invertible linear map between on the one hand the $G$-invariant subspace of
the quantum Hilbert space over $M$ and on the other hand the quantum Hilbert
space over $M^{G}.$ They interpret this result as a form of quantization
commuting with reduction.

However, Guillemin and Sternberg do not show that this invertible linear map
is unitary, and indeed there seems to be no reason that it should be in
general. So in their setting we may say that quantization fails to commute
\textit{unitarily} with reduction. Dan Freed \cite{F} has suggested to me that
inclusion of the half-form correction in the quantization might the map
unitary, and indeed our Yang-Mills example seems to confirm this. (It was
Freed's suggestion that led me to work out that $u$ is just the half-form
correction.) After all, upon inclusion of the half-form correction we get the
same measure (except for an irrelevant overall constant) and therefore the
same inner product whether quantizing before or after the reduction.
Nevertheless, I do not believe that one will get a unitary correspondence in
general, even with half-forms.

We are left, then, with the following open question.

\begin{quote}
Given a K\"{a}hler manifold $M$ with an action of a group $G,$ under what
conditions on $M$ and $G$ will quantization commute \textit{unitarily} with reduction?
\end{quote}

Although the question may be considered with or without the half-form
correction, what little evidence there is so far suggests that the answer is
more likely to be yes if the half-form correction is included.

\section{Notes}

\textit{Section 2}. One should say something about the degree of smoothness
assumed on the connections and gauge transformations. Although it does not
matter so much at the classical level, it seems natural to take the space of
connections to be the Hilbert space of square-integrable connections. This
amounts to completing $\mathcal{A}$ with respect to the natural norm, the one
which appears in the formula for the classical Hamiltonian. We may then take
the gauge group to be the largest group whose action on $\mathcal{A}$ makes
sense. This is the group of ``finite-energy'' gauge transformations, namely,
the ones for which $\left\|  g^{-1}\,dg\right\|  $ is finite. It is easily
shown that in our example of a spatial circle, two square-integrable
connections are related by a finite-energy based gauge transformation if and
only if they have the same holonomy. In the quantized theory we will be forced
to consider a larger space of connections.

The geodesics in $K$ (in the reduced dynamics) are relative to the
bi-invariant Riemannian metric determined by the chosen Ad-invariant inner
product on the Lie algebra.

\textit{Section 3}. The measure $P_{s}$ is a Gaussian measure, about which
there is an extensive theory. For example, see \cite{G1,Ku,GJ}. The
distinctive feature of Gaussian measures on infinite-dimensional spaces is the
presence of two different spaces, a Hilbert space $H$ whose norm enters the
formal expression for the measure, and a larger topological vector space $B$
on which the measure lives. Although one should think of the Gaussian measure
as being canonically associated to $H,$ the measure lives on $B,$ and $H$ is a
measure-zero subspace. In our example $H$ is the space of square-integrable
connections and $B$ is a suitable space of distributional connections. Since
the elements of $B$ are highly non-smooth, the holonomy must be defined as the
solution of a \textit{stochastic} differential equation.

If one glosses over questions of smoothness, the Gross ergodicity theorem
\cite{G2} sounds trivial. But we have just said that we must enlarge the space
of connections in order for the measure $P_{s}$ to exist. Unfortunately, we
may not correspondingly enlarge the gauge group without losing the
quasi-invariance of the measure $P_{s}$ under the action of $\mathcal{G}_{0},$
without which the definition of $L^{2}\left(  \mathcal{A},P_{s}\right)
^{\mathcal{G}_{0}}$ does not make sense. In the end two connections with the
same holonomy need not be $\mathcal{G}_{0}$-equivalent, because the would-be
gauge transformation is not smooth enough to be in $\mathcal{G}_{0}.$ It was
the ``J-perp'' theorem, which arose as a corollary of Gross's proof of the
ergodicity theorem, which led him to suggest to me to look for an analog of
the Segal--Bargmann transform on $K.$

\textit{Section 4}. Driver and I define the holomorphic subspace of
$L^{2}\left(  \mathcal{A}_{\mathbb{C}},M_{s,\hbar}\right)  $ to be the $L^{2}$
closure of the space of holomorphic cylinder functions. An important question
then is whether a function of the form $F\left(  Z\right)  =\Phi\left(
h_{\mathbb{C}}\left(  Z\right)  \right)  ,$ with $\Phi\in\mathcal{H}%
L^{2}\left(  K_{\mathbb{C}},\mu_{s,\hbar}\right)  ,$ is in this holomorphic
subspace. The answer is yes, but the proof that we give is indirect.

I am \textit{defining} $\mathcal{H}L^{2}\left(  \mathcal{A}_{\mathbb{C}%
},M_{s,\hbar}\right)  ^{\mathcal{G}_{0}}$ to be the image of $L^{2}\left(
\mathcal{A},P_{s}\right)  ^{\mathcal{G}_{0}}$ under the Segal--Bargmann
transform. Certainly every element of $\mathcal{H}L^{2}\left(  \mathcal{A}%
_{\mathbb{C}},M_{s,\hbar}\right)  ^{\mathcal{G}_{0}}$ is actual invariant
under the action of $\mathcal{G}_{0}$ on $\mathcal{A}_{\mathbb{C}}.$ The
converse is probably true as well, namely that every element of $\mathcal{H}%
L^{2}\left(  \mathcal{A}_{\mathbb{C}},M_{s,\hbar}\right)  $ which is
$\mathcal{G}_{0}$-invariant is in the image of $L^{2}\left(  \mathcal{A}%
,P_{s}\right)  ^{\mathcal{G}_{0}},$ but we have not proved this.

\textit{Section 5}. Except when $s=$ $\hbar$ the coherent states $\psi_{Z}$ in
$L^{2}\left(  \mathcal{A},P_{s}\right)  $ are non-normalizable states. When
$s=$ $\hbar,$ the coherent states $\psi_{Z}$ are normalizable states provided
that $Z$ is a square-integrable (complex) connection \cite[Sect. 2.3]{HS}. But
even then the measure $M_{\hbar,\hbar}$ does not live on the space of
square-integrable connections, and so it is a bit delicate to formulate the
resolution of the identity. This shows that it is technically easier to
formulate things in terms of the Segal--Bargmann transform instead of the
coherent states. Nevertheless, we may continue to think of unitarity for the
Segal--Bargmann transform as formally equivalent to a resolution of the
identity for the coherent states.

\textit{Section 6}. There are several obstructions to (\ref{mod.gc}) holding
in general. One needs some condition to guarantee that the analytic
continuation of the $G$-action exists globally. Even when it does, one needs
to worry about the possibility of ``unstable points,'' that is points whose
$G_{\mathbb{C}}$-orbit does not intersect the zero set of the moment mapping,
and also about the possibility that the $G_{\mathbb{C}}$-orbits may not be
closed. In the case of a cotangent bundle of a compact Riemannian manifold
whose cotangent bundle admits a global adapted complex structure, none of
these problems actually arises. See \cite[Sect. 7]{A}.

\textit{Section 8. }I jumping to conclusions about the correct action of the
gauge group $\mathcal{G}_{0}$ on the Segal--Bargmann space $\mathcal{H}%
L^{2}(\mathcal{A}_{\mathbb{C}},M_{s,\hbar}).$ One should properly use
geometric quantization to determine this action. To do this, we restrict first
to the finite-dimensional space $\mathcal{H}L^{2}(\mathbb{C}^{d},\nu_{\hbar})$
and then consider the action of the group of rotations and translations of
$\mathbb{R}^{d}$ on this space. Going through the calculations, on finds that
with my normalization these rotations and translations act in the obvious way,
namely, by composing a function in $\mathcal{H}L^{2}(\mathbb{C}^{d},\nu
_{\hbar})$ with the rotation or translation. Note that this holds only for
rotations and translations in the $x$-directions. Now we have said that the
action of $\mathcal{G}_{0}$ on $\mathcal{A}$ consists just of a rotation and a
translation. So, taking $\mathcal{H}L^{2}(\mathcal{A}_{\mathbb{C}},M_{s,\hbar
})$ as the best approximation of $\mathcal{H}L^{2}(\mathbb{C}^{d},\nu_{\hbar})
$ when $d=\infty,$ it is reasonable to say that the action of $\mathcal{G}%
_{0}$ on $\mathcal{H}L^{2}(\mathcal{A}_{\mathbb{C}},M_{s,\hbar})$ should be
just $F\left(  Z\right)  \rightarrow F\left(  g^{-1}\cdot Z\right)  .$

There is a large body of work extending the results of \cite{GStern}; see for
example the survey article of Sjamaar \cite{Sj}.

\end{document}